\documentclass{aastex63}

\shorttitle{}
\shortauthors{}

\newcommand{\Msun}{M$_{\odot}$}

\newcommand{\Lsun}{L$_{\odot}$}

\newcommand{\M}{M$_{BD}$\ }

\newcommand{\Neii}{[Ne\,{\scriptsize II}]}
\newcommand{\OIa}{[O\,{\scriptsize I}]\,$\lambda$6300}

\newcommand{\SIIa}{[S\,{\scriptsize II}]\,$\lambda$6731}

\newcommand{\F}{[FeII] 1.644$\mu$m\ }

\begin{document}

\title{Optical Forbidden Emission Line Spectro-astrometry of T~CrA: Evidence for a Multiple System and Multiple Jets \footnote{Based on data collected by UVES (089.C-0299(A) observations at the VLT on Cerro Paranal (Chile) which is
operated by the European Southern Observatory (ESO).}}

\author{Whelan, E.T.}
\affiliation{Maynooth University Department of Experimental Physics, National University of Ireland Maynooth, Maynooth Co. Kildare, Ireland}
\author{Murphy, A.}
\affiliation{Maynooth University Department of Experimental Physics, National University of Ireland Maynooth, Maynooth Co. Kildare, Ireland}
\author{Pascucci, I.}
\affiliation{Lunar and Planetary Laboratory, The University of Arizona,
Tucson, AZ 85721, USA}


\begin{abstract}
Spectro-astrometry is applied to echelle spectra of the young intermediate mass star T CrA. The aim is to better understand the origin of the [O I] and [S II] emission from T~CrA and further explore the usefulness of spectro-astrometry to the search for a reliable tracer of MHD disk winds. The analysis reveals a small-scale curved jet in an East-West direction and inclined parallel to the plane of the sky. It is the inclination of this jet which led to the classification of the forbidden emission lines as a low-velocity component. Thus spectro-astrometry highlights here that for close to edge-on disks spatial information is necessary. The position angle of the jet is not perpendicular to the position angle of the accretion disk nor does it agree with older observations of outflows likely driven by T~CrA. The mass outflow rate of 5 - 10 $\times$ 10$^{-8}$ \Msun/yr is within the range for intermediate mass stars. We conclude that more than one outflow is driven by the T CrA system and that the curvature seen in the first detection of an outflow from T CrA and in the data presented here is likely due to the multiplicity of the system.

\end{abstract}


\section{Introduction} 

Young, low to intermediate mass stars and brown dwarfs, have all been found to be accreting and to be surrounded by accretion disks of gas and dust \citep[e.g.,][]{Hartmann2016}  In addition, these accretors have spectra bright in forbidden emission lines (FELs) \citep[e.g.,][]{Whelan2014a}. Early spectroscopic studies of young stars revealed these lines to have multiple velocity components, with the high velocity component (HVC) tracing a jet and the origin of the low velocity component (LVC) less certain \citep{Hirth1997}. A velocity component with a peak velocity typically $<$ 30~kms$^{-1}$ is defined as a LVC while a HVC has a peak velocity $>$ 100~km$s^{-1}$. Later studies showed that these components could be further decomposed, with individual sources having more than one HVC and with the LVCs having a narrow and a broad component (NC, BC) \citep{Simon2016, Banzatti2019, Whelan2021}. In recent years, interest in the FEL LVC has intensified due to its significance to the question of angular momentum removal from the  disks of young accreting objects \citep[e.g., the recent review by ][]{Pascucci2022}. 

For several decades, magneto-rotational instability (MRI)-induced turbulence and magneto-hydrodynamic (MHD) winds have been the competing theories dominating thinking on how angular momentum is removed from disks and accretion at the observed rates enabled \citep{Pascucci2022}. Recent
simulations find that non-ideal MHD effects suppress MRI
over a large range of disk radii ($\sim$ 1-30~au), in support of radially extended MHD disk winds as the prime means for angular momentum extraction in disks \citep{Pascucci2022, Lesur2022}. To drive this theory forward a reliable observational tracer of disk winds must be identified and this is where the FEL LVC becomes significant \citep{Ercolano2017, Pascucci2020}. In particular, the LVC of the \OIa\ emission line has been suggested as a tracer of MHD winds and several high resolution spectroscopic studies have used kinematical studies of this and other FELs (e.g. SII]$\lambda$4068, \SIIa\ [Ne II] 12.81~$\mu$m), to test the robustness of the FEL LVC as a MHD disk wind tracer \citep{Simon2016, Fang2018, Banzatti2019, Sacco2012}. These studies generally support the origin of the LVC BC in an MHD wind but results are less clear for the LVC NC. 

One weakness of such high resolution kinematical investigations is that they do not offer any information on the spatial properties of these typically compact LVC FEL regions. This is where spectro-astrometry comes to the fore. Spectro-astrometry when applied to high spectral resolution data can offer kinematical and spatial information which can be used to uncover the origin of the FEL regions. This was demonstrated in \cite{Whelan2021} where spectro-astrometry was used to show that the \OIa\ and \SIIa\ LVC NC was likely tracing an MHD disk wind for the case of the young low-mass star RU~Lupi. For the young star AS~205~N the analysis supported a complicated multiple system with multiple jets. 
Examples of how this technique has been used previously include identifying unresolved stellar jets and stellar companions \citep{Whelan2008} and detecting the rotation of accretion disks \citep{Pontoppidan2011}. Here we apply the technique of \cite{Whelan2021} to the intermediate mass young star T Coronae Australis (hereafter T~CrA) with the goal of better understanding the origin of its FELs and further investigating the usefulness of spectro-astrometry to the study of MHD winds.


\begin{figure*}
\centering
   \includegraphics[width=18cm,trim= 0cm 6cm 0cm 2cm, clip=true]{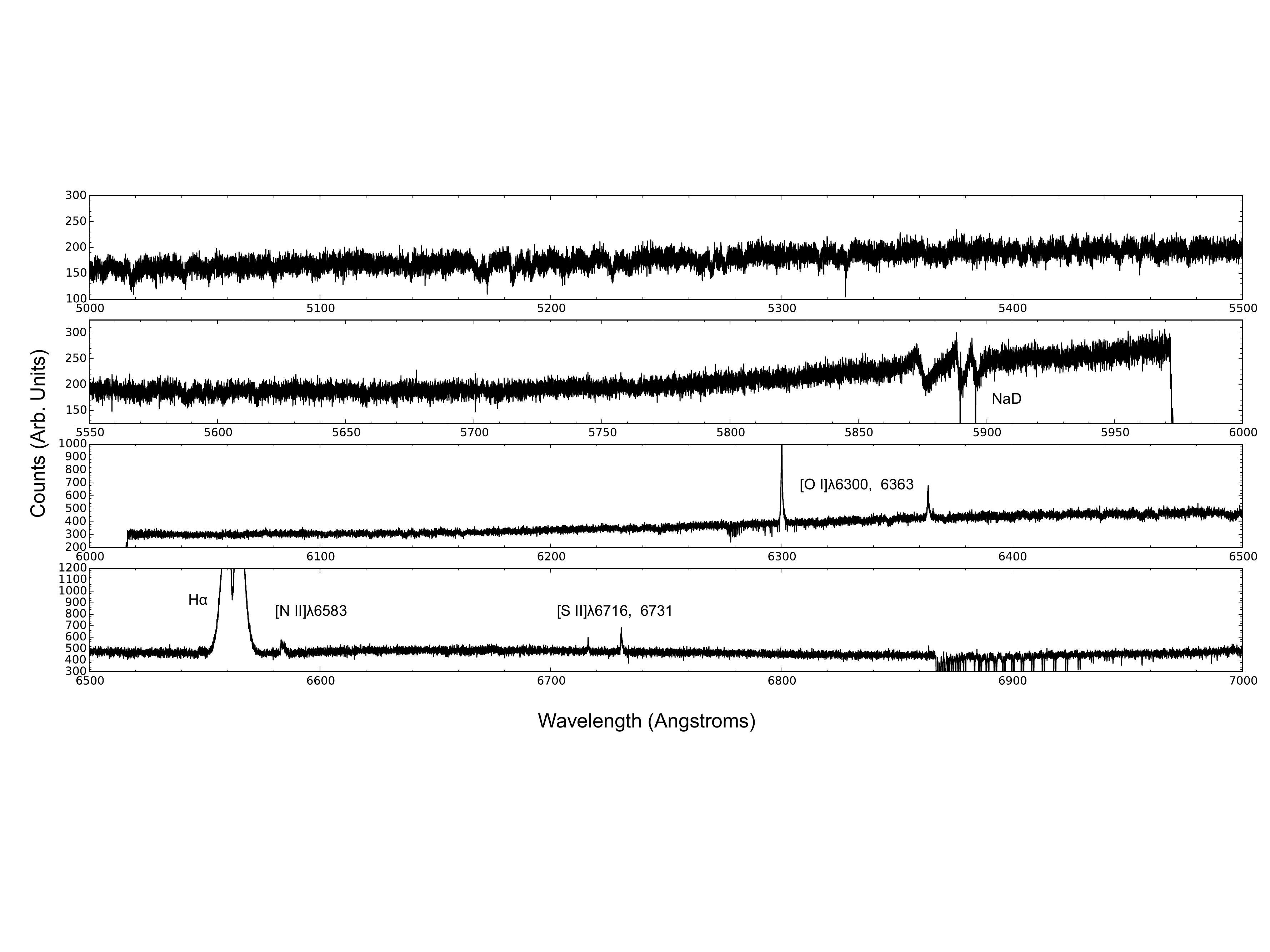}
     \caption{Full UVES spectrum of T CrA with the jet tracing lines marked.}
  \label{TCRA_SPEC}     
\end{figure*}

T~CrA is an intermediate mass young star located in Corona Australis at d$\sim$ 150~pc \citep{Galli2020}. Relevant properties of T~CrA are given in Table \ref{tprops}. Studies of the SED of T CrA indicate the presence of a large primordial circumstellar disk extending outward to a radius greater than 100 AU \citep{Sicilia-Aguilar2013}. \cite{Cazzoletti2019} give the position angle (PA) of the disk at 20$^{\circ}$ using ALMA observations. Recent observations by \cite{Rigliaco2023} using SPHERE/IRDIS put the disk PA at $\sim$ 7$^{\circ}$ and support an edge-on inclination for the disk. The detection of Herbig-Haro (HH) objects around T~CrA indicate that it drives an outflow \citep{Hartigan1987}. It is also thought to be a binary star system with spectro-astrometry constraining the companion separation at $\approx$ 260 mas and the binary PA at 275$^{\circ}$ \citep{Bailey1998, Takami2003}. The binary detection comes from spectro-astrometric analysis of the H$\alpha$ line. This is further supported by the analysis of the light curve of T~CrA by \cite{Rigliaco2023}. They conclude that the T~CrA binary has a period of 29.6 years and constrain the binary PA at 130$^{\circ}$ $\pm$ 15$^{\circ}$ (310$^{\circ}$ $\pm$ 15$^{\circ}$). A detection of a slow wind is reported in \cite{Sacco2012} based on a small blue-shift in the \Neii{} line. \cite{Pontoppidan2011} investigated the CO emissions from T~CrA with spectro-astrometry and classified it as a self-absorbed source where the absorbing gas is structured on much larger scales than the emitting gas.

The FELs of T~CrA are single peaked with velocities close to 0~kms$^{-1}$ thus they are classified as LVCs.  A high-resolution spectrum covering the \OIa\ was recently reported in \cite{Pascucci2020}, see their fig.~10. The line is decomposed into two components both of which have a small velocity shift with respect to the stellar velocity, and hence could be interpreted as LVC (their Table 5). Here we probe with spectro-astrometry the \OIa\ and \SIIa\ FELs in detail for the first time. The observations and the observing strategy particular to spectro-astrometry are discussed in Section 2. The spectro-astrometric study is presented in Section 3 and in Section 4 what new information we have gleaned on this system and on the usefulness of spectro-astrometry to the question of the origin of the FEL LVC is discussed. 

\begin{figure*}
\centering
   \includegraphics[width=20cm]{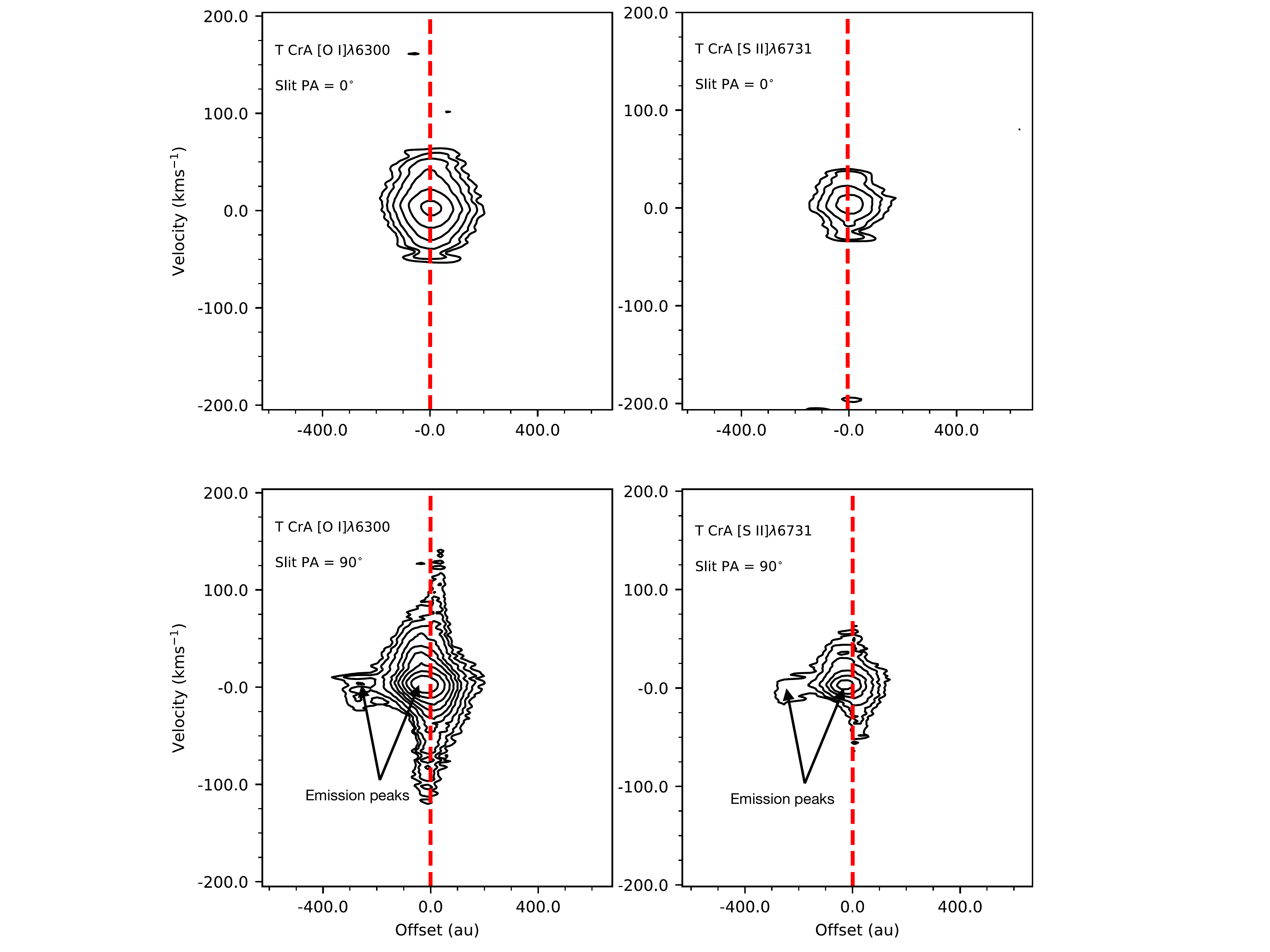}
     \caption{Position velocity diagrams of the \OIa\ and \SIIa\ lines at 0$^{\circ}$ and 90$^{\circ}$. The emission peaks at 0~kms$^{-1}$ but it is clearly extended at 90$^{\circ}$ with extended line wings.}
  \label{TCRA_PV}     
\end{figure*}

\begin{figure*}
\centering
   \includegraphics[width=7.5cm]{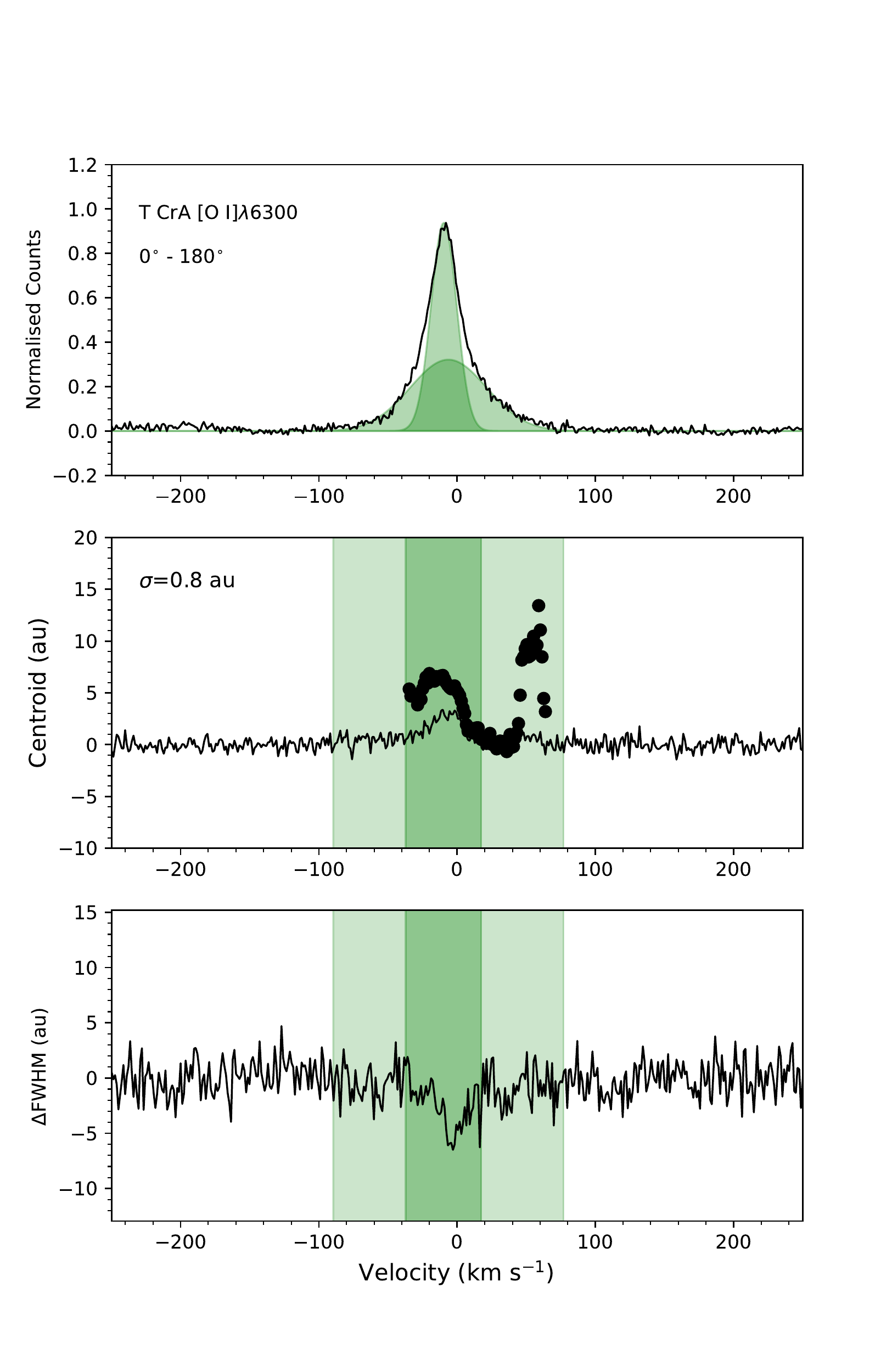}
    \includegraphics[width=7.5cm]{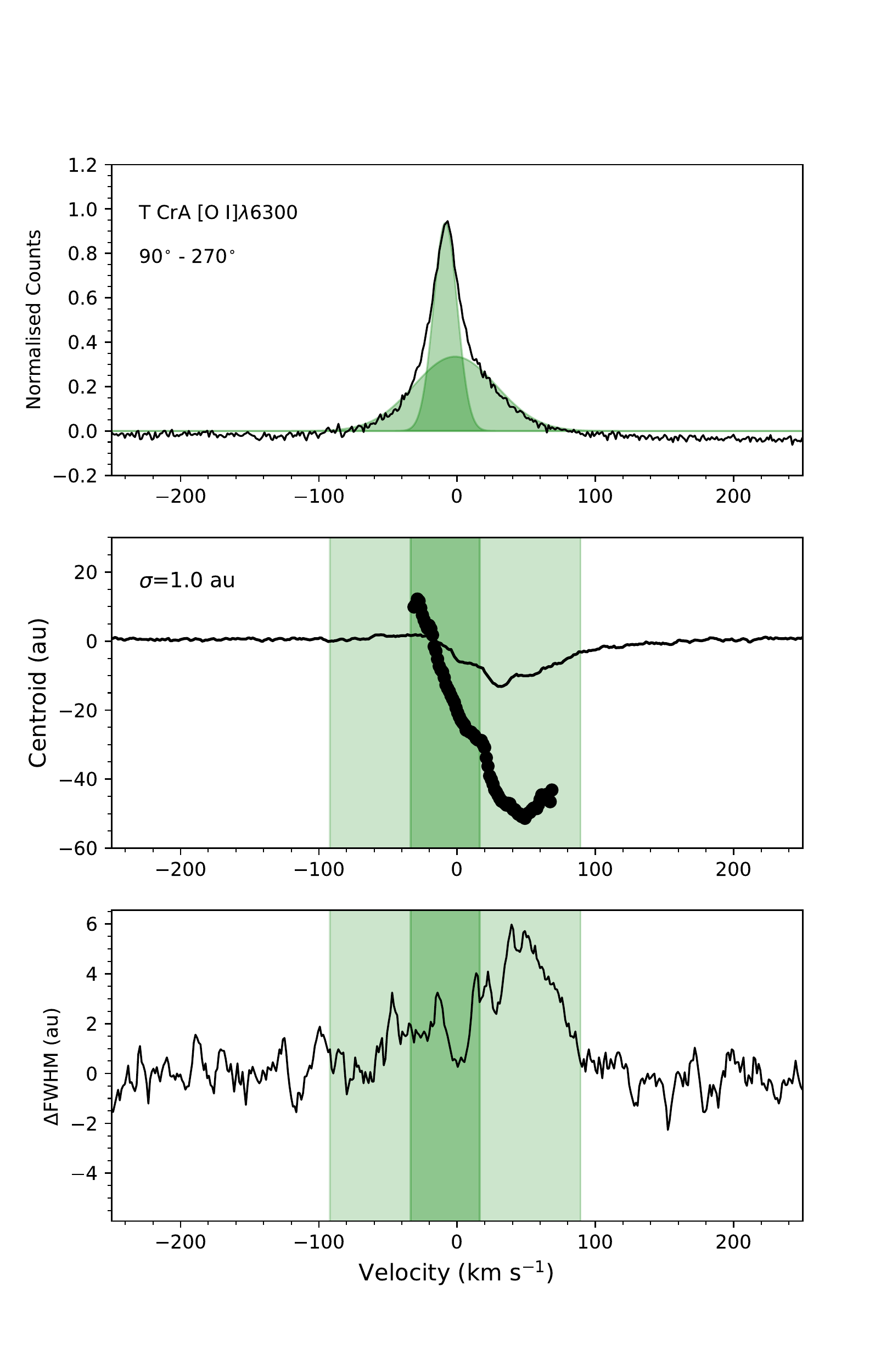}
     \includegraphics[width=7.5cm]{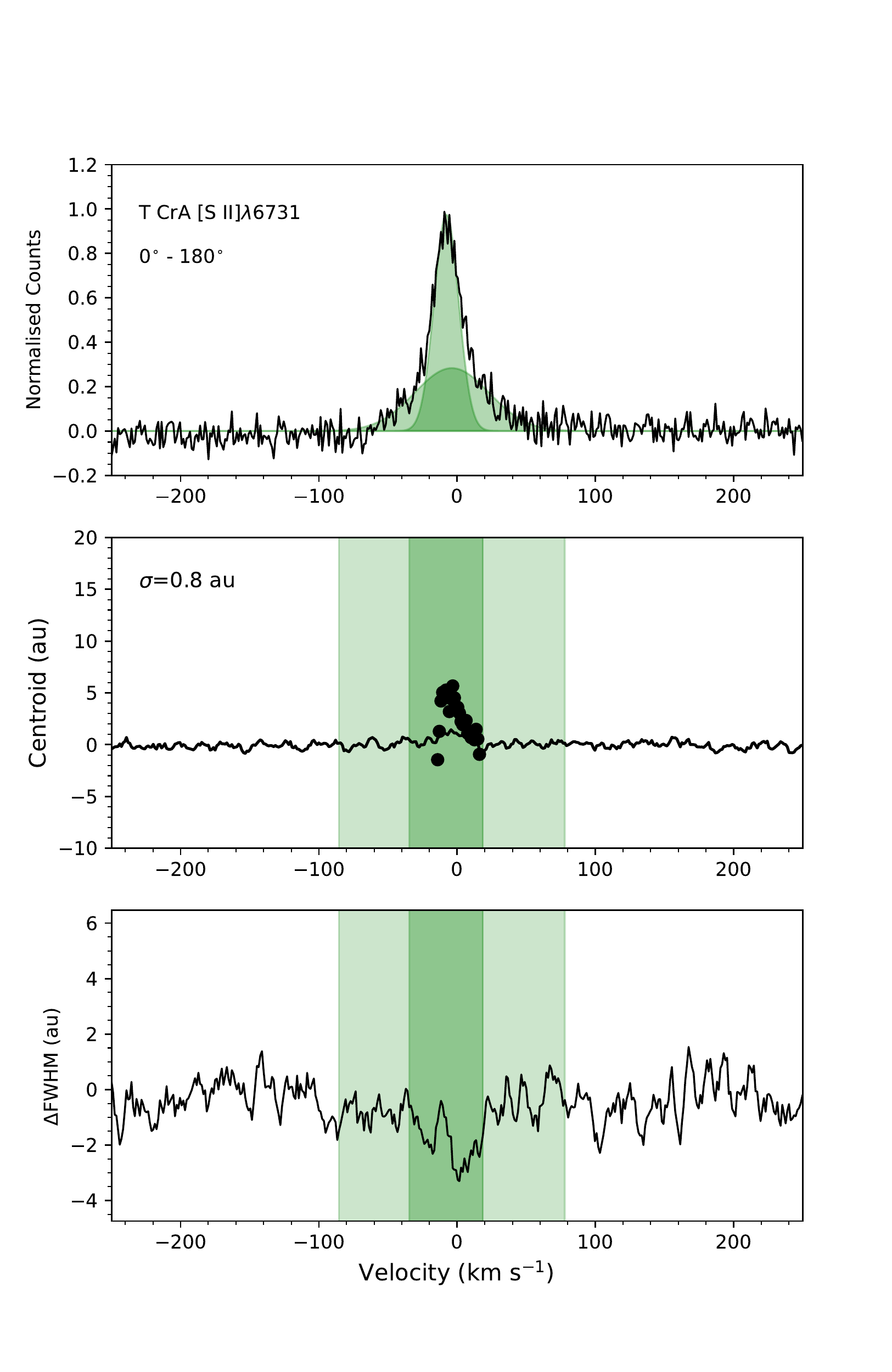}
    \includegraphics[width=7.5cm]{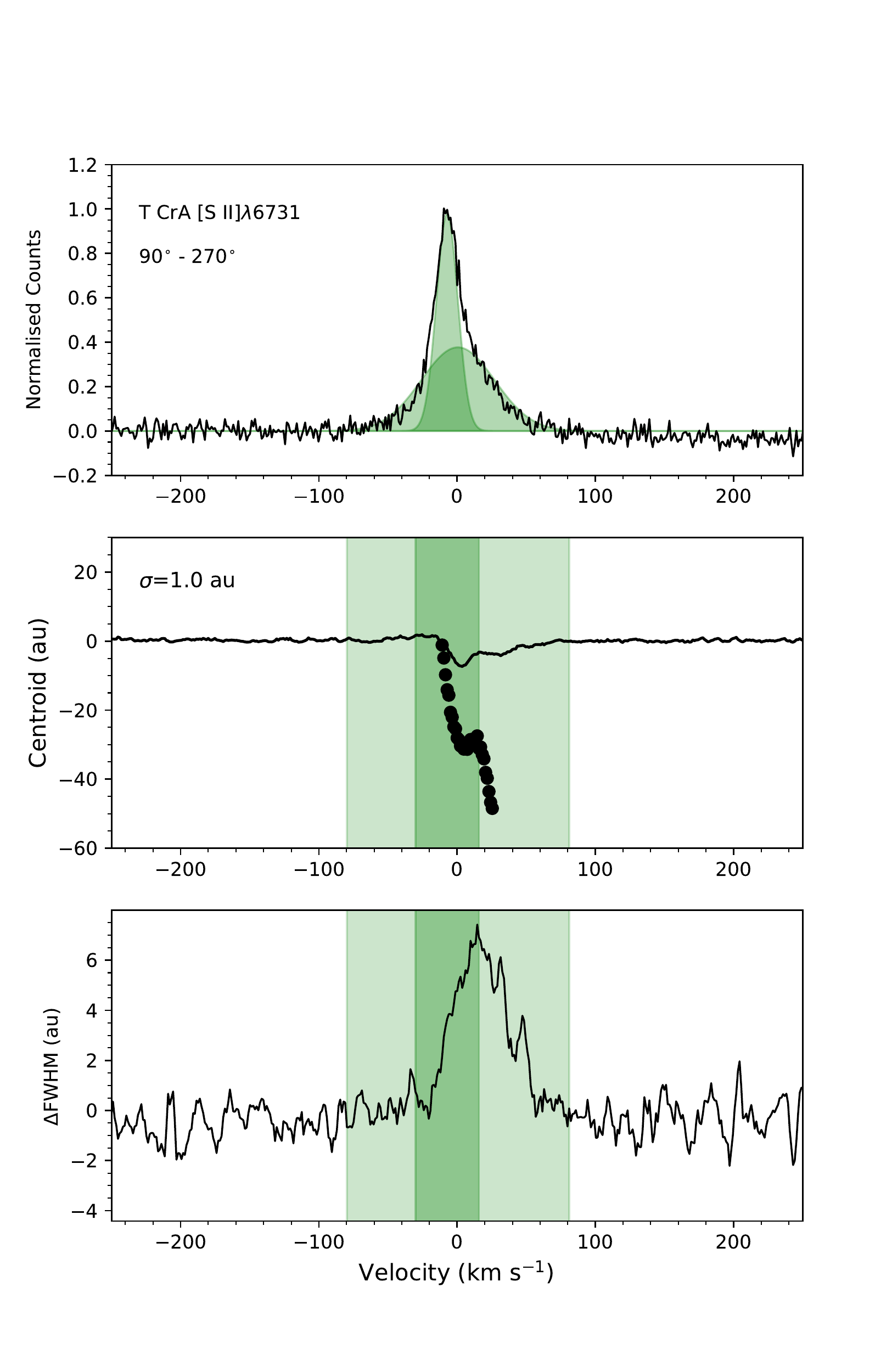}
     \caption{{\bf T~CrA:} Spectro-astrometric analysis of the [O I]$\lambda$6300, and [S II]$\lambda$6731 lines. Top: The continuum subtracted line emission, normalised to the line peak. Middle: The position spectra with the ranges of the velocity components shown as the colored shaded regions. The continuum subtracted offsets are over-plotted in black. Bottom: The intrinsic widths of the emission regions.}
     \label{TCRA_SA}     
   \end{figure*}

\section{Observations and Spectro-astrometry}
Spectra of T~CrA were collected with the UV-Visual Echelle Spectrograph (UVES) on the European Southern Observatory's Very Large Telescope (ESO VLT) \citep{Dekker00} in June 2012 through the ESO programme 089.C-0299(B) (P.I. Pascucci). The slit width was set to 1\arcsec{} and a spectral resolution of R = 40,000 was achieved over a wavelength range of $\sim$ 5000~\AA\ to 7000~\AA{}. The slit was rotated to four position angles (PAs) starting with the PA of the accretion disk and rotating by 90$^{\circ}$ in each case (see Table 2). In this way spectra were obtained parallel and anti-parallel to the accretion disk  (0$^{\circ}$ and 180$^{\circ}$) and parallel and anti-parallel to any outflow emission (90$^{\circ}$ and 270$^{\circ}$), assuming that the PA of the outflow emission is perpendicular to the disk PA. As discussed below this strategy is integral to the spectro-astrometric analysis. At the time of the observations the estimate of the PA of the T CrA disk was 0$^{\circ}$ hence the slit PAs were 0$^{\circ}$, 90$^{\circ}$, 180$^{\circ}$ and 270$^{\circ}$. As noted above the PA of the T CrA accretion disk has since been estimated using ALMA data at 20$^{\circ}$. This does not change the effectiveness of the technique. 

The data were reduced using the ESO pipeline and standard IRAF routines. The spectro-astrometric process involves fitting the spatial profile of a spectrum in order to determine the spatial centroid of the emission. This is done at regular intervals across the wavelength range\footnote{Typically every pixel or a bin of a few pixels can be used to increase the SNR and thus accuracy of the fitting} to produce what is known as a a position spectrum. In general Gaussian fitting is used and the accuracy to which the centroid can be measured depends on the SNR of the spectrum \citep{Porter2004}. The centroid of emission line regions will differ from the centroid of the continuum emission if the emission lines are tracing an outflow for example or are excited by one component of a binary \citep{Mendigutia2018}. The centroid measurements are a lower limit on the positions of the emission regions as they are weighted by the position of the brightest emission.

The reasoning behind the strategy of taking spectra at four slit PAs is two fold. Firstly, the anti-parallel slit PAs are used to remove any artefacts which can be introduced \citep{Brannigan2006} and the perpendicular slit PAs can be used to construct a 2D map of the emission region under analysis \citep{Whelan2012}. Finally in order to recover an accurate spatial profile of any emission region contamination by the continuum emission should be corrected for \citep{Whelan2008}. 

\begin{deluxetable}{lll}
\tablecaption{Properties of T CrA \label{propsample}}
\tablehead{
 \colhead{Property}
 &  & \colhead{References}
 }
\startdata
SpTy   &F0  &C19 \\
Mass (M$_\odot$) &1.8 &D18\\
Log($\dot{M}_{\rm acc}$) (M$_\odot$/yr)  & -7.94 &P20 \\
Disk Type   &full &C19 \\
Disk PA ($^\circ$)  &20 &C19\\
\enddata
\tablecomments{References stand for: C19 = Cazzolettei et al. (2019); D18 = Dong et al. (2018), GL06 = Garcia-Lopez et al. (2006)} 
\label{tprops}
\end{deluxetable}

\begin{deluxetable}{ccc} 
\tablecaption{Observing log \label{sample}}
\tablehead{
 \colhead{Date} & \colhead{Slit PA} & \colhead{Exposure Times} \\
  \colhead{[dd-mm-yyyy]} & \colhead{[$^\circ$]} & \colhead{[ncycle$\times$s]}
}
\startdata
 29-06-2012 &0 &5 $\times$ 450  \\ 
 29-06-2012 &90 &5 $\times$ 450  \\ 
29-06-2012 &180 &5 $\times$ 450  \\ 
29-06-2012 &270 &5 $\times$ 450 \\
\enddata
\label{obs}
\end{deluxetable}

\begin{deluxetable}{cccc} 
\tablecaption{Kinematical Fitting \label{sample}}
\tablehead{
 \colhead{Line} & \colhead{Slit PA} & \colhead{V$_{c}$} & \colhead{FWHM}  \\
   & \colhead{[$^\circ$]} & \colhead{km~s$^{-1}$} & \colhead{km~s$^{-1}$}
}
\startdata
\OIa  &0 &-10 &22 \\ 
      & &-6 &65 \\
\OIa  &90 &-9 &20 \\ 
      & &-2 &71 \\
\SIIa &0 &-8 &21 \\ 
 & &-4 &64 \\
\SIIa &90 &-7 &18 \\
 & &1 &63 \\
\enddata
\label{kin}
\end{deluxetable}


\section{Results}

\begin{figure*}
\centering
   \includegraphics[width=18cm, width=18cm,trim= 0cm 5cm 0cm 5cm, clip=true]{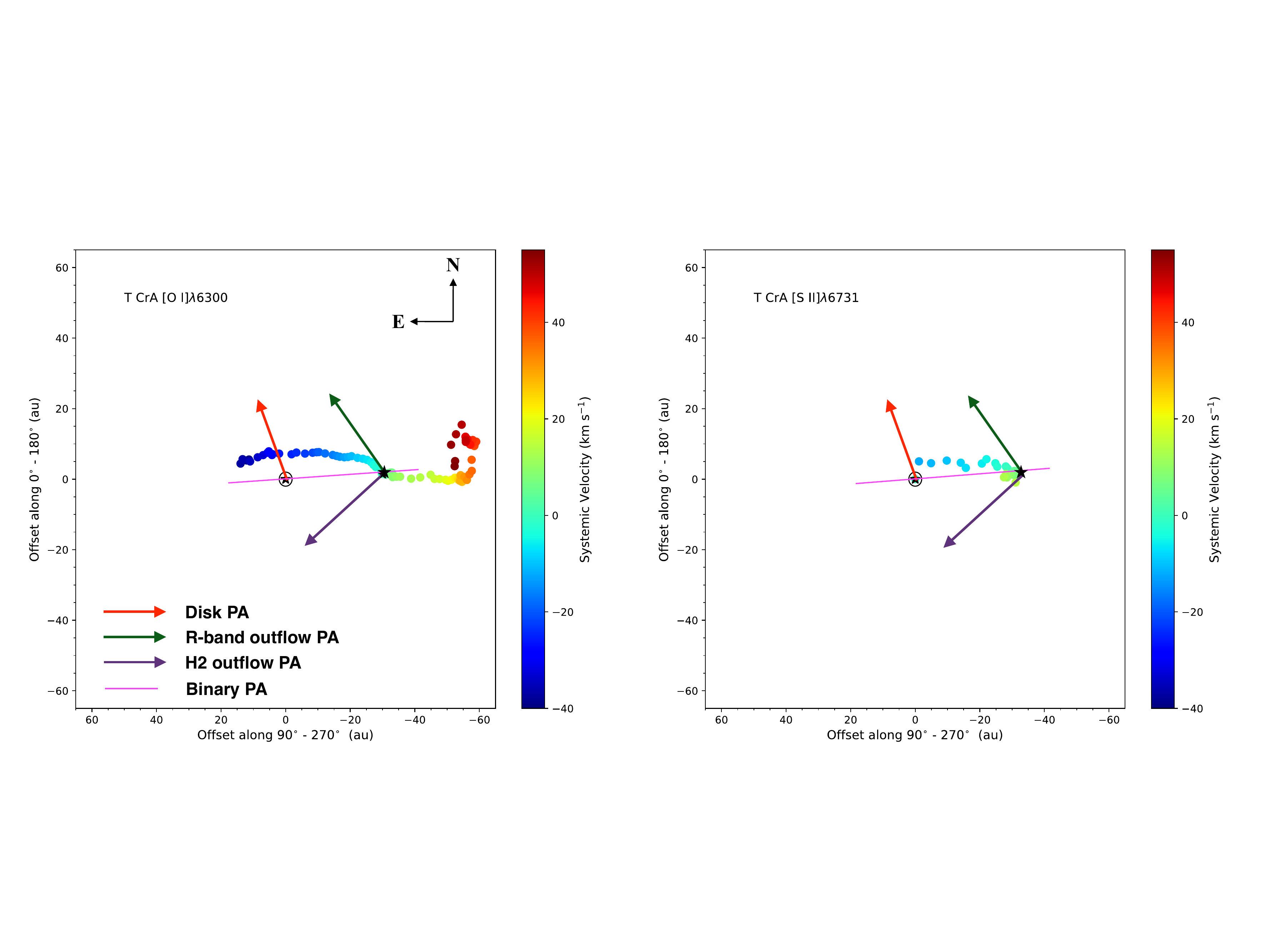}
     \caption{{\bf T~CrA:} Combining the perpendicular slit PAs to recover the 2D spatial information. The continuum emission has been subtracted here to reveal the full offset in the outflow. The black circle is the 3-$\sigma$ error in the centroid. The green star is the centroid of the continuum emission. The black star is the position at which the jet emission changes from blue-shifted to red-shifted. The colored lines represent the PAs described in the legend.}
  \label{TCRA_SA_2D}     
\end{figure*}

The spectral range of the UVES data covers numerous emission lines tracing outflow and accretion. In the case of T~CrA only the H$\alpha$, [O I]$\lambda$6300, 6363 [N II]$\lambda$6583 and [S II]$\lambda\lambda$6716, 6731 emission lines are detected (Figure \ref{TCRA_SPEC}). For the purpose of this spectro-astrometric study we focus on the two brightest FELs namely the \OIa\ and \SIIa\ lines. Note that the [O I]$\lambda6363$ line shows the same results as \OIa\ line and the [N II]$\lambda$6583, [S II]$\lambda$6716 lines as the \SIIa\ line. All velocities quoted are in the stellocentric reference frame, and distances in Figures \ref{TCRA_PV}, \ref{TCRA_SA} and \ref{TCRA_SA_2D} have not been corrected for the inclination of the system with respect to the plane of the sky. In Figure \ref{TCRA_PV} continuum subtracted position velocity (PV) diagrams of the \OIa\ and \SIIa\ emission regions at slit PAs of 0$^{\circ}$ and 90$^{\circ}$ are presented. At 0$^{\circ}$ the emission is coincident with the stellar position (0~au) while at 90$^{\circ}$ two emission peaks at $\sim$ -50~au and $\sim$ -350~au are seen. This comparison between the two slit PAs reveals an outflow at a PA of close to 90$^{\circ}$. The radial velocities of the emission peaks are close to 0~kms$^{-1}$.


Figures \ref{TCRA_SA} and \ref{TCRA_SA_2D} present the results of the spectro-astrometric analysis of the emission regions shown in Figure \ref{TCRA_PV}. In Figure \ref{TCRA_SA} the slit PAs along (left column) and perpendicular (right column) to the assumed accretion disk PA are compared. There are four panels with each panel consisting of three plots. The top two panels give the results for the \OIa\ line and the bottom two for the \SIIa\ line. The top row of each panel shows the emission line profiles. The line profiles are single peaked with a red-shifted wing. They are best fit with two velocity components and the radial velocities of the components for the \OIa\ and \SIIa\ lines are given in Table 3. From considering the spectro-astrometry in combination with the PV plots we conclude that the bulk of the emission is tracing a jet hence we identify these components as HVCs are therefore they are coloured green following previous colour schemes \citep{Whelan2021}.

The middle row of each panel of Figure \ref{TCRA_SA} shows the position spectra. The anti-parallel slit PAs have been subtracted to remove any artefacts. The 1-$\sigma$ accuracy in the continuum centroid is given. As spectro-astrometry only gives the emission centroid, what is shown here for the 90$^{\circ}$/270$^{\circ}$ PAs is the position of the emission peak closest to the source, as it is the brightest emission. This corresponds to the outflow knot at $\sim$ 50~au in the 90$^{\circ}$ PV plot. The outflow knot at $\sim$ 350~au has a negligible effect on the measured centroid as it is much fainter. The centroids of the continuum subtracted emission regions are over-plotted as black dots. In some cases the subtraction of the continuum increases the displacement of the emission region with respect to the continuum, by an amount equal to the ratio of the brightness of the (continuum + line) emission to the line emission \cite{Takami2001}. In other cases e.g. the \SIIa\ line at the slit PA of 0$^{\circ}$/180$^{\circ}$, a displacement in the line region is not revealed until after the continuum is removed. This is due to the faintness of the pure line emission relative to the continuum in this case. In the bottom row of all four panels of Figure \ref{TCRA_SA} the change in the FWHM of the emission region as a function of velocity is plotted. This provides information on the extent of the emitting region. 

In Figure \ref{TCRA_SA_2D}, the continuum subtracted centroids plotted in the left and right columns of Figure \ref{TCRA_SA}, are combined into a 2D picture and plotted as a function of systemic velocity. The red arrow shows the PA of the disk taken from the ALMA observations of \cite{Cazzoletti2019}. The green and purple arrows mark the PAs of known outflows detected from optical and infrared imaging \citep{Ward1985, Kumar2011}. These outflows are discussed in detail in Section 5. Finally the magenta line delineates the PA of the detected companion. From Figures \ref{TCRA_SA} and \ref{TCRA_SA_2D} it is clear that the TCrA system is more complicated than it appears in Figure \ref{TCRA_PV} i.e. more complicated than a single star driving a jet.  In combination Figures \ref{TCRA_SA} and \ref{TCRA_SA_2D} tell us two things about the T~CrA system, which are now described. 

\subsection{A Low Inclination Misaligned Jet}

Figures \ref{TCRA_SA} and \ref{TCRA_SA_2D} reveal that the FEL region of T~CrA is tracing a jet inclined close to the plane of the sky. This inclination is inferred from the fact that the red-shifted jet emission is traced back to the source position and the radial velocity of the lines close to 0~kms$^{-1}$. This inclination minimises the radial velocity of the jet tracing lines and leads to the LVC classification for the FELs.  This conclusion is supported by the increase in the FWHM of the FEL regions at slit PAs 90$^{\circ}$/270$^{\circ}$ This is of course expected for outflow emission. The inclination of the jet is expected for a star with an edge-on disk \citep{Rigliaco2023}. Also note from Figure \ref{TCRA_SA_2D} that the displacement of the emission away from the star in the jet, generally increases with velocity, which is a signature of a magneto-hydrodynamically launched outflow \citep{Whelan2021}. Figure \ref{TCRA_SA_2D} shows the PA of jet to be be close to 90$^{\circ}$ which is not exactly perpendicular to the disk PA of 7$^{\circ}$ to 20$^{\circ}$, although it does suggest that the disk PA is closer to 7$^{\circ}$ than 20$^{\circ}$. Furthermore, the blue shifted jet initially curves away from the source with a PA closer to 45$^{\circ}$. This is discussed below and adds uncertainty to any measurement of the jet PA.

\subsection{Multiple Companions}

Figure \ref{TCRA_SA_2D} also reveals the presence of at least one companion to T~CrA. 
The centroid of the continuum emission is the (0,0) point (marked with a green star). If the continuum centroid corresponds to the position of the driving source of the jet then the outflow emission would switch from being red-shifted to blue-shifted (and change direction) at this position. This is not the case for T~CrA and the position at which this does happen is marked by the black star. The black star corresponds to the position of the jet emission at 0~kms$^{-1}$. Hence, the continuum emission is tracing two sources and the continuum centroid is the intermediate position between the two sources which depends on the their relative brightness \citep{Bailey1998}. The black star is the position of the source which is driving the jet and therefore a line between the the black and green stars gives the binary PA and a lower limit to the binary separation.

This conclusion is again supported by the intrinsic width measurements of Figure \ref{TCRA_SA}. The intrinsic width decreases for both the \OIa\ and \SIIa\ lines, in the left column. In the case of unresolved companions this can be due to only one component of the binary emitting in the line under investigation \citep{Mendigutia2018}. \cite{Bailey1998, Takami2003} used the same analysis of the H$\alpha$ emission from T~CrA to detect a companion to T~CrA with a PA of 275$^{\circ}$ and a minimum separation of 20~au to 40~au. The same companion is being detected here in the FELs with a separation of at least 30~au and a PA of $\sim$ 274$^{\circ}$ (marked with a magenta line). Something new which this FEL analysis reveals is a variation or curving in the jet axis. The \OIa\ and \SIIa 2D spectro-astrometry in Figure \ref{TCRA_SA_2D} shows that the blue-shifted jet is curved within the first $\sim$10~au of the driving source and that the red and blue jets curve away from each other.
There is a hint of this in the H$\alpha$ spectro-astrometry presented in \cite{Takami2003} although it is hard to be certain as the red and blue-shifted velocities are not separated in their Figure 3. This change in the position of the jet axis with distance from the driving star can be explained by a further companion interacting with the circumstellar disk \citep{Murphy2021} (see Section 4). This curvature also add uncertainty to the estimate of the binary PA using spectro-astrometry and likely accounts for the difference with the PA estimate of \cite{Rigliaco2023}. 

\section{Discussion and Conclusions}

The first reference to an outflow driven by T~CrA is made by \cite{Ward1985} who identified a dusty disk surrounding the source from polarisation measurements and a jet at a PA of 130$^{\circ}$ to 135$^{\circ}$ from R band imaging. They report that the jet is curved towards the south and that the jet and disk are misaligned with the angle between the jet and disk being $\sim$ 60$^{\circ}$. They estimate the projected length of the jet to be 0.012~pc. \cite{Hartigan1985} studied a larger region around T~CrA, again with R Band imaging and mark the extended emission identified by \cite{Ward1985} as feature {\it k} in their Figure 13. Again this faint jet feature appears curved. \cite{Rigliaco2023} use SPHERE/IRDIS to image the region around T ~CrA and identify extended emission which is likely the same emission first imaged by \cite{Ward1985, Hartigan1985}. They argue however that this feature is not a jet but an accretion streamer tracing infalling material.
\cite{Hartigan1987} carried out further R band imaging of T~CrA and identify HH~98 along an approximate E-W PA and at a distance of $\sim$ 0.04~pc from T~CrA. This HH object is slightly red-shifted in agreement with results presented here (Figure 4). \cite{Wang2004} conclude from [S II] imaging that the HH object HH~733 is most likely driven by T~CrA. This flow would have a PA of $\sim$ 35$^{\circ}$. \cite{Kumar2011} studied the H$_{2}$ flows around T~CrA and identified a H$_{2}$ counterpart to HH~733 labelled MHO~2013. They also identify MHO~2015 to the South of T~CrA which is clearly the counter flow to HH~733/MHO~2013. The PA of the full H$_{2}$ flow is also 35$^{\circ}$. This flow is again detected at a PA of 35$^{\circ}$ by \cite{Rigliaco2023}. 

Almost 40 years of studies of the region around T~CrA show the outflow activity to be complicated. From our review of the literature it is concluded that two significant outflows have been identified to date. The first is approximately E-W (HH~98) which we clearly see on small scales with spectro-astrometry and with some curvature. The second flow observed at optical and infrared wavelengths has a PA of 35$^{\circ}$. The first outflow is likely perpendicular to the edge-on disk although there could me a misalignment of a few degrees. This kind of misalignment has been detected previously for similar objects \citep{Murphy2021}. 
No clear evidence of this second flow outflow at the PA of 35$^{\circ}$ is identified in our UVES data but it is possible that the curvature in the blue-shifted jet at a PA of 45$^{\circ}$ seen in Figure 4 could be a signature of this second flow. Also the circular feature seen in Figure 4 at high red-shifted velocities for the \OIa\ line and in an approximate N-S direction could be caused by the presence of two outflows. AO corrected imaging is needed to disentangle the outflows and to constrain the number of jet driving sources in the system.

\begin{figure*}
\centering
   \includegraphics[width=12cm, trim= 0cm 1cm 0cm 0cm, clip=true]{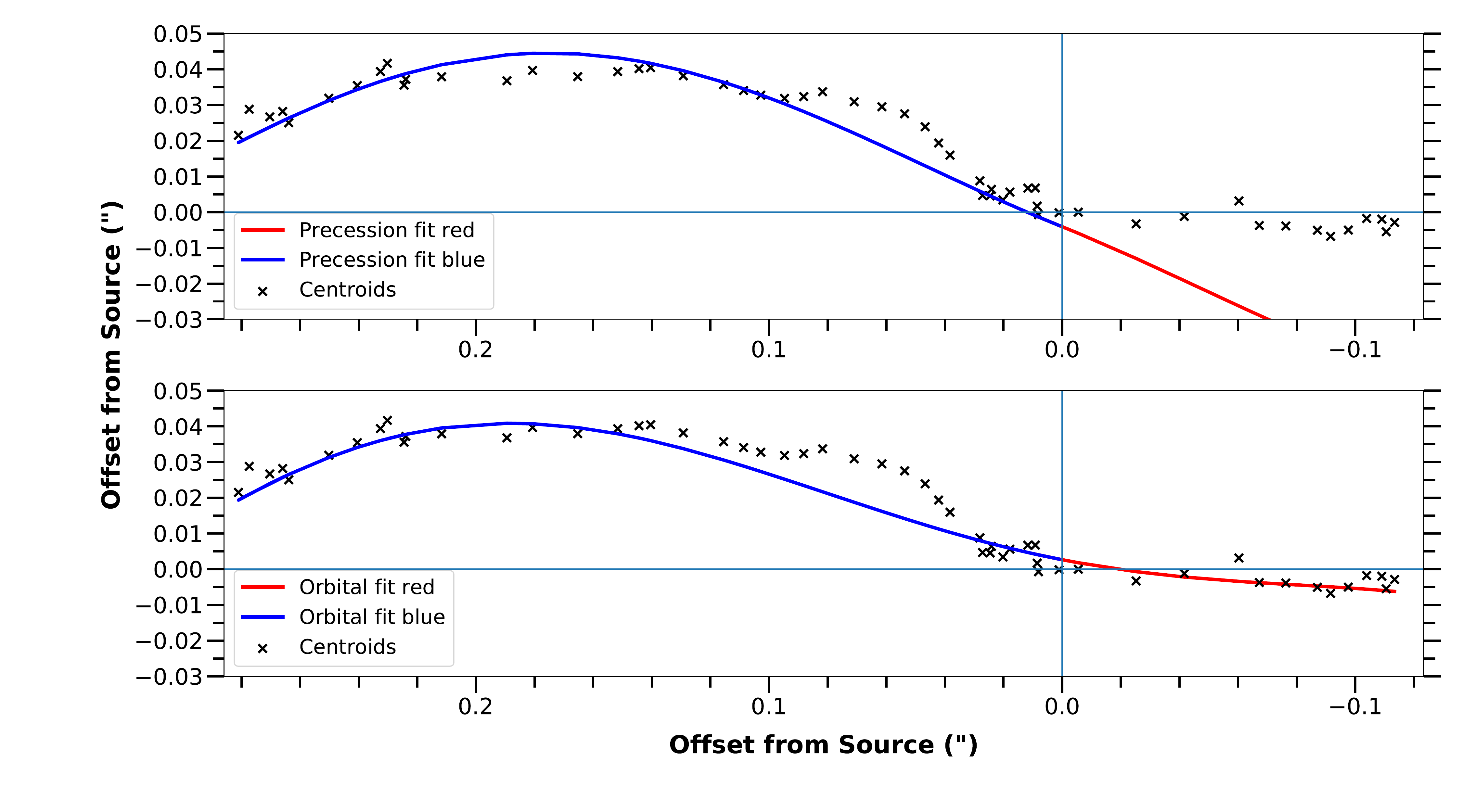}
     \caption{Estimated centroid positions for the T CrA jet, overlaid with best-fit models for precession (top) and orbital motion (bottom). The models for the blue- and red-shifted jet lobes are shown in the corresponding colors.}
  \label{T_CRA_wiggle}     
\end{figure*}

Evidence of curvature in the E-W outflow is clear from the spectro-astrometry presented here. The misalignment of both outflows with the disk could also be a signature of curved or wiggling jets. We also consider now that the curvature in the E-W flow could be caused by the presence of a third source. Following the method of \citet{Murphy2021}, the centroid positions of the more extended blue-shifted jet lobe were fitted with models for orbital motion and precession as described by \citet{Masciadri2002}. The best-fit models in each case are shown in Figure \ref{T_CRA_wiggle} overlaid on the jet centroids for both lobes. This analysis is based on the assumption of a low inclination for the jet, i.e. with the jet close to the plane of the sky. While both models can be fitted to the blue-shifted lobe alone, we see that the precession model matches poorly to the positions of the red-shifted lobe whereas the orbital motion model provides a more reasonable fit to the overall jet shape. This model does not take into account the effect of the outflow at a PA of 35$^{\circ}$ on the centroids and indeed for both models the initial portion of the blue lobe is not fit well. Assuming a typical jet velocity of 150~kms$^{-1}$, the orbital motion model suggests a 1.6~\Msun\ companion orbiting at a distance of 2.3~au. At much faster jet velocities (200~kms$^{-1}$) the companion mass becomes greater than the assumed total mass because this increases the required orbital velocity. Conversely, lower jet velocities imply a smaller companion at a slightly wider separation. Given the mass of T~CrA it is more likely that the velocity is $<$ 150~kms$^{-1}$. Therefore the chances that an undetected companion is causing the curvature in the blue lobe are very small, as we would expect to have detected such a companion already.

A further insight into the driving source of the E-W jet detected here can be gleaned from an estimate of the mass outflow rate ($\dot{M}_{\rm out}$). The luminosity of the \OIa\ line is estimated from the equivalent width (EW) at log~L$_{OI}$ = -3.36~\Lsun\ \citep{Whelan2009}. Following the method of \cite{Fang2018} and assuming a velocity range of 200~kms$^{-1}$ to -400~kms$^{-1}$ $\dot{M}_{\rm out}$ is estimated at 5-10 $\times$ 10$^{-8}$~\Msun~yr$^{-1}$. This estimate is in the range of values found for objects with a similar spectral type and mass as T~CrA \citep{Nisini1995}. Combining $\dot{M}_{\rm out}$ with the mass accretion rate given in \cite{Pascucci2020} ($\dot{M}_{\rm acc}$ = 5-10 $\times$ 10$^{-8}$~\Msun~yr$^{-1}$) gives a jet efficiency of 5$\%$ to 9$\%$. Efficiencies of $<$ 10$\%$ are typically found for Classical T Tauri stars and Herbig Ae/Be stars \citep{Frank2014,Pascucci2022}. This points to the primary of the system with a spectral type of F0 and a mass of 1.8~\Msun\ as the driver of the E-W jet. The results presented in Figure \ref{TCRA_SA_2D} do not rule out the possibility that the companion identified here and by \cite{Bailey1998, Takami2001} as the driver of the almost N-S jet. This source lies somewhere along the magenta line drawn in Figure 4 and with a minimum separation of $\sim$ 30~au from the driver of the E-W jet (marked with a black star). This conclusion is supported by the results of \cite{Rigliaco2023}
which suggest that the circumbinary disk is perpendicular to the N-S jet.

It is concluded that there are at least two sources within the T~CrA system driving at least two outflows. This kind of multiplicity is common and has been seen before in Z CMa and AS~205~N for example \citep{Whelan2010, Whelan2021}. The most significant result from this piece of work is that it demonstrates that spectro-astrometry, which gives kinematical and spatial information at the same time, is the most efficient way of separating low and high forbidden emissions from young stars whose disks are seen close to edge-on and for investigating the origin of the emission. As spectro-astrometry gives a lower limit on the extent of the flows follow-up high angular resolution observations with JWST or VLT/MUSE would be very effective for mapping the full extent of the flows.


{}

\end{document}